\documentclass[aps,prl,showpacs,twocolumn]{revtex4}
\usepackage{graphicx,epsfig}
\begin{document}
\title{Information transfer and nontrivial collective behavior\\ in
chaotic coupled map networks}
\author{L. Cisneros}
\affiliation{ Laboratorio de Fen\'onemos no Lineales, Escuela de F\'{\i}sica,
Facultad de Ciencias\\ Universidad Central de Venezuela, Caracas, Venezuela.}
\author{M. G. Cosenza}
\affiliation{Centro de Astrof\'{\i}sica Te\'orica, Facultad de Ciencias,
Universidad de Los Andes, \\Apartado Postal 26 La~Hechicera, M\'erida~5251,
Venezuela.}
\author{J. Jim\'enez}
\affiliation{ Laboratorio de Fen\'onemos no Lineales, Escuela de F\'{\i}sica,
Facultad de Ciencias\\ Universidad Central de Venezuela, Caracas, Venezuela.}
\author{A. Parravano}
\affiliation{Centro de Astrof\'{\i}sica Te\'orica, Facultad de Ciencias,
Universidad de Los Andes, \\Apartado Postal 26 La~Hechicera, M\'erida~5251,
Venezuela.}

\begin{abstract}
The emergence of nontrivial collective behavior in networks of coupled chaotic
maps is investigated by means of a nonlinear mutual prediction method. 
The resulting prediction error is used to measure the amount of information that a local
unit possesses about the collective dynamics.
Applications to locally and globally coupled map systems are considered.
The prediction error exhibits
phase transitions at critical values of the coupling for the onset of ordered
collective behavior in these networks. This information measure may be used 
as an order parameter for the characterization of
complex behavior in extended chaotic systems.
\end{abstract}
\date{Accepted in Phys. Rev. E, Rapid Communications (2002)}
\maketitle
Much
interest has recently been directed to understanding the phenomenon of emergence of
nontrivial collective behavior in systems of interacting chaotic elements
\cite{Chate}. Nontrivial collective behavior (NTCB) is characterized by a well-defined
evolution of macroscopic quantities emerging out of local chaos.
Models based on coupled map networks have been widely used in the investigation of
cooperative phenomena that appear in many extended chaotic dynamical systems \cite{Chaos}. 
In particular, NTCB has been
studied in coupled maps on regular Euclidean lattices \cite{Chate,Mann}, in
one-dimensional lattices \cite{Co1}, fractal geometries \cite{Co2}, and globally
coupled map systems \cite{Kaneko,Pikovsky,Shibata,Gallego}. In this article, we
investigate the information transfer between the local and global levels of coupled
map network models as a condition for self-organization in spatiotemporal systems. One
may ask the question: how much information does a local unit possess about the
collective dynamics of a system?; or, how does the information flow depend on
parameters of the system?.

To analyze how global order in spatiotemporal systems can arise out of
local chaos, we consider 
a system of $N$ interacting elements, where  the state of element $i$ $(i=1,2,\ldots,N)$
at discrete time $t$ is denoted by $x_t^i$. The evolution of each element is assumed to
depend
on its own local map dynamics and on its interaction with other elements in the network,
where the strength of the interactions is given by a coupling parameter.
The collective dynamics of the system at time $t$ may be described by some statistical 
variable $h_t$, such as the mean field.
The information transfer between local ($x_t^i$) and macroscopic ($h_t$)
variables is analyzed by comparing their corresponding time evolutions. We adopt a simple
computational technique based on a mutual nonlinear modeling
\cite{Schiffetal,Anrhold}. This method makes use of the temporal 
evolution of a driven
variable (which is receiving information) to infer characteristics of the driver
variable. In our case, for different values of the coupling parameter we record time series
containing the simultaneous evolution of the macroscopic quantity $h_t$ and of a chosen local
variable $x^i_t$.

The trajectory of the series $x^i_t$ is reconstructed in an embedding space of
dimension $d$ as a collection of vectors $(x^i_t, x^i_{t-1}, \dots, x^i_{t-d+1})$.
Then, for each of these vectors, we systematically look for its nearest neighbor
$(x^i_{p}, x^i_{p-1}, \dots, x^i_{p-d+1})$ in the Euclidean distance, as shown in
Fig.~1. The root-mean-square error over $h_t$ at the embedding dimension $d$ is
computed as
\begin{equation}\label{error}
  E_d(h|x)= \frac{\langle (h_{t+1}-{h}_{p+1})^2 \rangle^{1/2}}{\sigma} \, ,
\end{equation}
where $h_{p}$ is the value of the macroscopic variable that bears the time index of $x^i_{p}$,
and $\sigma$ is the standard deviation of $h_t$.

\begin{figure}[ht]
\centerline{\hbox{
\epsfig{file=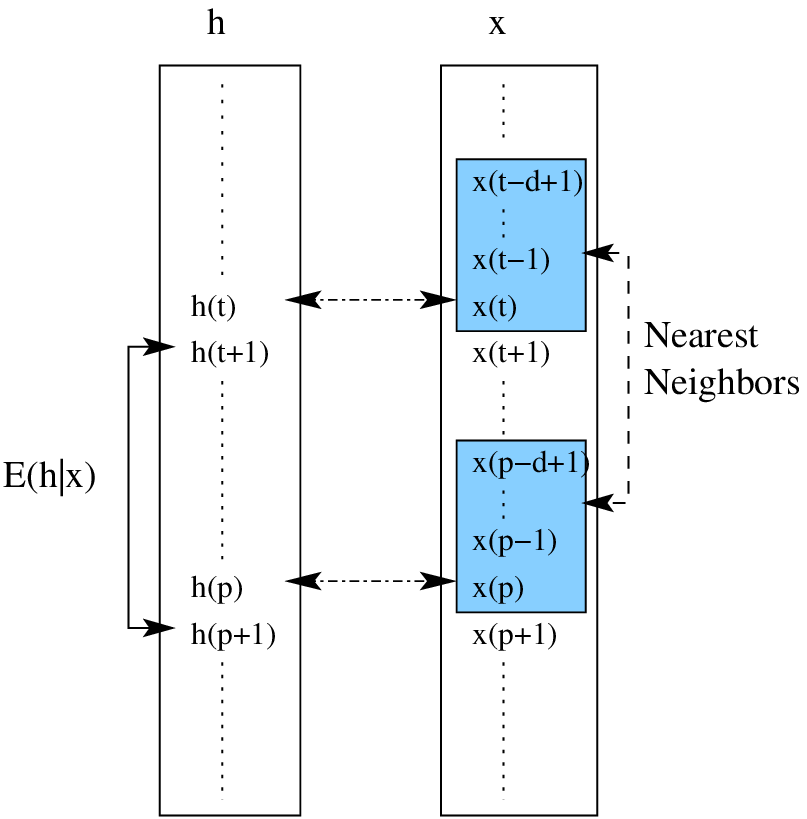, width=.40\textwidth,clip=}
}}
\caption{(a) Bifurcation diagram $h_t$ vs. $\epsilon$ for homogeneous globally
coupled maps, with $b=0$. Four different phases are observed as $\epsilon$ is
varied: turbulent (T), periodic (P), chaotic bands (C), and synchronized (S). (b)
$E_3(h|x)$ vs $\epsilon$ for this system. (c) Bifurcation diagram of a local map$x_t^i$ vs. $\epsilon$, exposing the underlying chaotic dynamics.}
\end{figure} 

The prediction error given by Eq.~(\ref{error}) can be interpreted as a 
measure of the information
that the series $x^i_t$ possesses about the macroscopic variable $h_t$. In this context, small
values of $E_d(h|x)$ imply that the dynamics of the global quantity $h_t$ is embedded
in the local evolution. Larger mutual prediction errors indicate that the two time
series become more independent of each other. In particular, for two totally
independent random series the error Eq. (\ref{error}) has a value of $\sqrt{2}$.

As applications, we have calculated  the error $E_d(h|x)$ as a function of the 
coupling strength in coupled map systems exhibiting NTCB.
This quantity was computed for several embedding dimensions
$d$ and the curve giving minimal errors was selected in each case.

The first example is a system of maps subjected to global coupling defined as
\begin{equation}\label{hgcm}
 x_{t+1}^i= (1-\epsilon)f_i(x_t^i)+\frac{\epsilon}{N}\sum_{j=1}^N  f_j(x_t^j),
\end{equation}
where the function $f_i(x_t^i)$ describes the local dynamics of element $i$, 
and $\epsilon$ is the coupling parameter. The usual homogeneous
globally coupled map system \cite{Kaneko} corresponds to having the same local
function for all the elements, i.e., $f_i(x_t^i)=f(x_t^i)$.
As local dynamics, we employ the logarithmic map $f(x)= b + \ln |x|$ \cite{Kawabe},
where $b$ is a real parameter. This map does not belong to the standard classes of
universality of unimodal or bounded maps.
Robust chaos occurs in the parameter
interval $b \in [-1,1]$, with no periodic windows and no separated chaotic bands on
this interval \cite{Kawabe}.
In a second example, heterogeneity in the local maps in Eq. (\ref{hgcm}) is
introduced by taking $f_i(x_t^i)= b_i + \ln |x_t^i|$, with the values of the
parameters $b_i$ distributed in $[-1,1]$.

The last example is a one-dimensional, homogeneous, diffusively coupled 
logarithmic map lattice given by
\begin{equation}\label{cml}
  x_{t+1}^i=(1- \epsilon)f(x^i_t) +
  \frac{\epsilon}{2}\left[f(x_{t}^{i+1})+f(x_{t}^{i-1})\right],
\end{equation}
where periodic boundary conditions are assumed in Eq. (\ref{cml}).

As the macroscopic variable for these systems
we consider the instantaneous mean field, defined as
\begin{equation}\label{mean}
  h_t=\frac{1}{N}\sum_{j=1}^N  f_j(x_t^j).
\end{equation}

\begin{figure}[ht]
\centerline{\hbox{
\epsfig{file=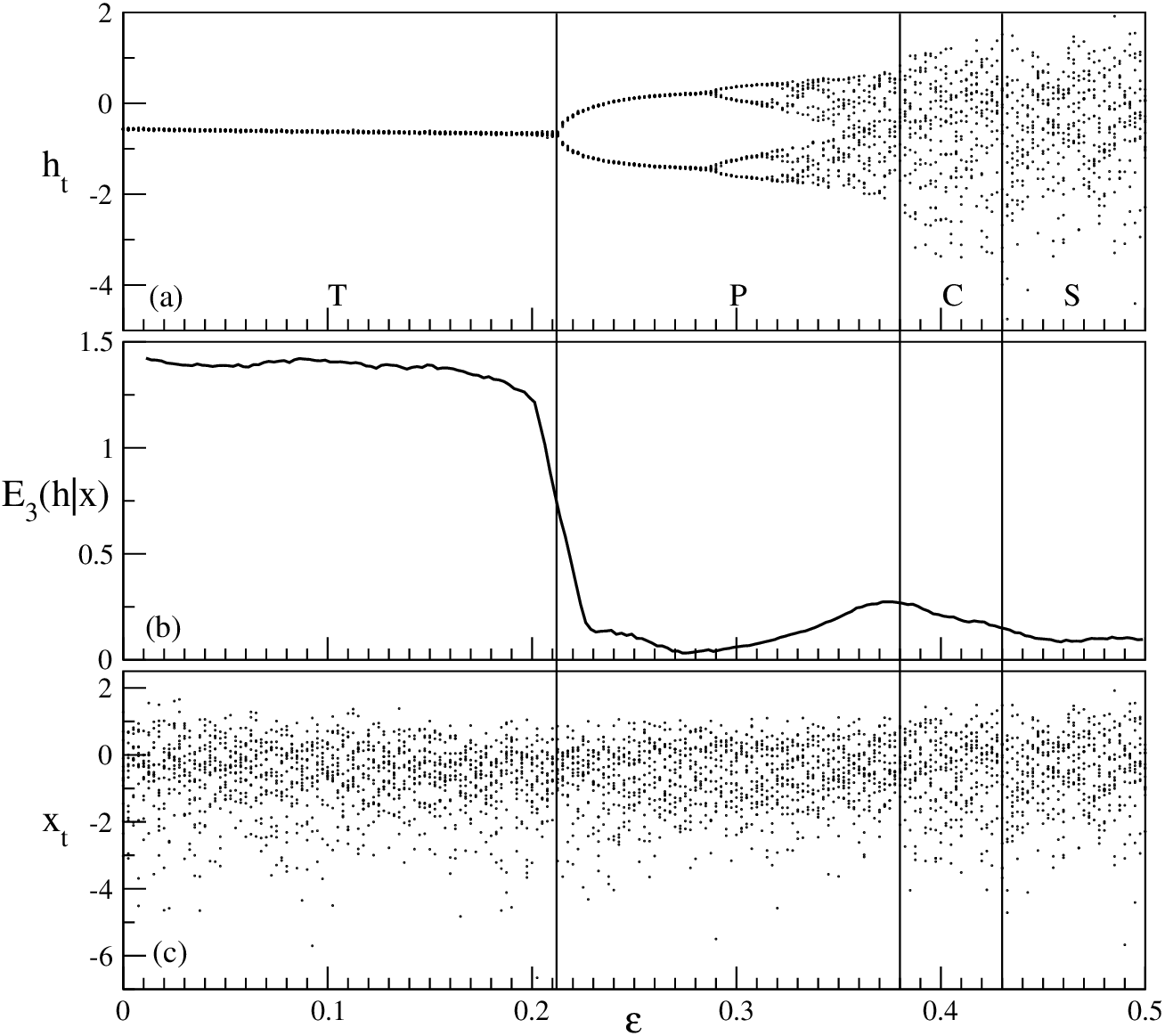, width=.40\textwidth,clip=}
}}
\caption{(a) Bifurcation diagram $h_t$ vs. $\epsilon$ for homogeneous globally
coupled maps, with $b=0$. Four different phases are observed as $\epsilon$ is
varied: turbulent (T), periodic (P), chaotic bands (C), and synchronized (S). (b)
$E_3(h|x)$ vs $\epsilon$ for this system. (c) Bifurcation diagram of a local map
$x_t^i$ vs. $\epsilon$, exposing the underlying chaotic dynamics.}
\end{figure} 

Each of the above examples presents nontrivial collective
behavior in some range of their parameters. Figure 2(a) shows the bifurcation diagram
of the mean field $h_t$ of the homogeneous globally coupled map system,
Eq.~(\ref{hgcm}), as a function of the coupling strength $\epsilon$ \cite{Gallego}.
The local parameter is fixed at $b=0$ for all maps and the system size is $N=10^4$.
For each value of $\epsilon$, the mean field was calculated at each time step during a
run starting from random initial conditions on the local maps, uniformly distributed
on the interval $[-8,4]$, after discarding the transients. When the local parameter
$b$ is in the range $[-1,1]$, the elements $x_t^i$ are chaotic and desynchronized (see
Fig. 2(c)). However, the mean field in Fig. 2(a) reveals the existence of global
periodic attractors for some intervals of the coupling.
Different collective states emerge as a function of the coupling $\epsilon$: a
turbulent phase, where $h_t$ manifests itself as a fixed point, following the standard
statistical behavior of uncorrelated disordered variables; collective periodic states;
collective chaotic bands; and chaotic synchronization \cite{Gallego}. In this
representation, collective periodic states at a given value of the coupling $\epsilon$
appear as sets of vertical segments which correspond to intrinsic fluctuations of the
periodic orbits of the mean field. Increasing the system size $N$ does not decrease
the amplitude of the collective periodic orbits. Moreover, when $N$ is increased the
width of the segments that make a periodic orbit in the bifurcation diagrams such as
in Fig. 2(a) shrink, indicating that the global periodic attractors become better
defined in the large system limit.
Figure 2(b) shows
$E_3(h|x)$ vs. $\epsilon$. An abrupt change in the value of the quantity $E_3(h|x)$
can clearly be seen at a critical value of the coupling $\epsilon_c \simeq 0.21$, a
behavior characteristic of a first order phase transition. At this critical value of
the coupling the collective behavior changes from a turbulent phase to a periodic
collective state, as observed in Fig. 2(a). The error for $\epsilon<\epsilon_c$ is
large, indicating that there is no appreciable information sharing between the local
and macroscopic levels when the system is in the turbulent phase. For
$\epsilon>\epsilon_c$ the error drops discontinuously to very small values. The local
unit suddenly becomes ``aware" of the collective dynamics; the time series of a single
map is good enough to provide assertive predictions of the mean field evolution. Thus,
there is a large amount of information transfer from the collective dynamics to each
of the elements in the network, even before synchronization is achieved. Increasing
the coupling beyond the synchronization region leads again to a turbulent state of the
system and to a large value of the error $E_3(h|x)$.

\begin{figure}[ht]
\centerline{\hbox{
\epsfig{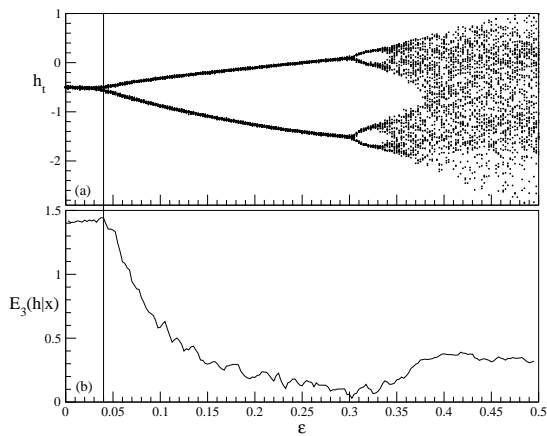}
}}
\caption{(a) Bifurcation diagram $h_t$ vs. $\epsilon$ for heterogeneous globally
coupled maps; system size is $N=10^4$. (b) $E_3(h|x)$ vs $\epsilon$ for this
network. }
\end{figure}

Figure 3(a) shows the bifurcation diagram of $h_t$ vs. $\epsilon$ for the globally
coupled heterogeneous map lattice. In this case the local parameters $b_i$ are set at
random with a uniform distribution in the chaotic interval, i.e., $b_i \in [-1,1]$.
Again the local dynamics are chaotic, yet collective periodic behavior arise in some
windows of the coupling parameter.
Figure 3(b) shows the error $E_3(h|x)$ vs. $\epsilon$ for the heterogeneous globally
coupled system. In this case the prediction error stays large up to a critical value
of the coupling $\epsilon_c \simeq 0.04$ and then decreases continuously for $\epsilon
> \epsilon_c$. The decrease in the prediction error resembles a second order phase
transition. The emergence of collective periodic behavior is manifested in the low
values of the error for $\epsilon > \epsilon_c$. The disorder introduced in the
globally coupled network by the local heterogeneity can be detected by the quantity
$E_3(h|x)$ as a change in the character of the transition to collective behavior when
compared to a similar transition in a homogeneous globally coupled system, Fig. 2(a).

The emergence of nontrivial collective behavior can also be observed in the
one-dimensional coupled map lattice, Eq. (\ref{cml}), as shown in Fig. 4(a). In
this case the system presents only a turbulent (statistical fixed point) phase
and a period-two collective state \cite{Co1}.
Figure 4(b) shows $E_2(h|x)$ as a function of the coupling in the one-dimensional
coupled map lattice. There is again a decrease in the error at the critical value of
the coupling strength for which the transition from turbulence to periodic collective
states occurs. We have observed similar transitions from large to small values in the
quantity $E_d(h|x)$ at the onset of NTCB in networks having other connecting
topologies, as well as when employing unimodal chaotic maps as local dynamics.

\begin{figure}[ht]
\centerline{\hbox{
\epsfig{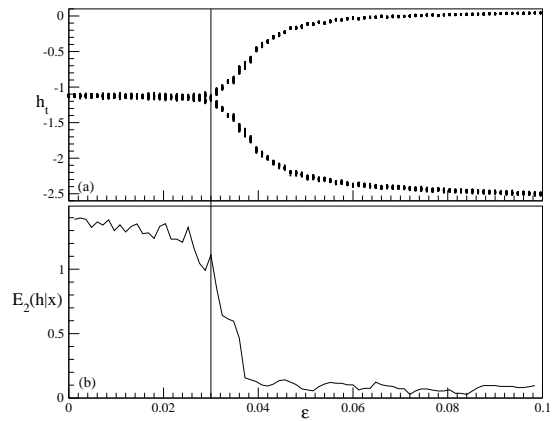}
}}
\caption{(a) Bifurcation diagram $h_t$ vs. $\epsilon$ for one-dimensional coupled map
lattice; $N=10^4$ and $b=0$. (b) $E_2(h|x)$ vs $\epsilon$ for this lattice. }
\end{figure}

The logarithmic map has been employed as local dynamics in the above examples
because the emergence of ordered collective behavior in those coupled systems
cannot be attributed to the existence of windows of periodicity nor to
chaotic band splitting in the local dynamics. These systems can be chaotic 
at a local level and simultaneously periodic at a macroscopic level. Thus, 
there should exist some 
global information sharing among the elements of the networks that leads to a
collective organization besides the trivial synchronization.
The information transfer required for the onset of nontrivial collective behavior 
takes place at some specific values of the parameters of the systems.  
The observed decreasing of the errors at the transition to nontrivial collective
behavior can be interpreted as a manifestation of the emergence of organization in
these systems. 
It should be noticed that in all cases the
dynamics of the elements do not experience notable change before and after the
transition to nontrivial collective behavior, since local dynamics is always chaotic.

In conclusion, we have shown that the error $E_d(h|x)$ is a useful quantity to
characterize the transition to ordered collective behavior in chaotic spatiotemporal
systems. Connectivity and coupling strengths are the mechanism for information flow in
networks of dynamical units. However, our results suggest that transference of the
information that is {\it relevant} for the emergence of collective organization in
systems of interacting chaotic elements is associated to low values of 
$E_d(h|x)$. Finally, the exploration of a possible relationship between 
$E_d(h|x)$ and other 
quantities used to study coupled chaotic map systems, such as the collective
Lyapunov exponent \cite{Shibata} or transfer entropy \cite{Schreiber},
is an interesting 
problem for future research. 

\section*{Acknowledgments}
M.G.C and A.P acknowledge support from Consejo de Desarrollo Cient\'{\i}fico,
Human\'{\i}stico y Tecnol\'ogico of Universidad de Los Andes, M\'erida, Venezuela. L.
Cisneros thanks Consejo Nacional de Investigaciones Cient\'{\i}ficas y Tecnol\'ogicas,
Venezuela, for support.

\end{document}